\documentclass[final, 5p, times, twocolumn, number, sort&compress]{elsarticle}

\usepackage[colorlinks=true,allcolors=blue]{hyperref}
\usepackage[scaled=0.85]{roboto-mono}
\usepackage[T1]{fontenc}
\usepackage{listings}
\usepackage{textcomp}
\usepackage{xcolor}
\usepackage{xparse}
\usepackage{graphicx}
\usepackage{caption}
\usepackage{upgreek}
\usepackage{dcolumn}
\usepackage{amsmath}
\usepackage{amsfonts}
\usepackage{amssymb}
\usepackage{mathtools}
\usepackage{bm}
\usepackage{cuted}
\usepackage{braket}
\usepackage{orcidlink}
\usepackage{nicefrac}

\usepackage{pythonhighlightcustom}

\newcommand{\eg}{\textit{e.g.}}
\newcommand{\etal}{\,\textit{et al.}}
\newcommand{\ie}{\textit{i.e.}}
\newcommand{\etc}{\textit{etc.}}

\newcommand{\qs}{\texttt{qspec}}
\newcommand{\qmod}{\texttt{qspec.models}}
\newcommand{\qsim}{\texttt{qspec.simulate}}
\newcommand{\qu}{\texttt{qutip}}
\newcommand{\st}{\texttt{satlas}}
\newcommand{\stt}{\texttt{satlas2}}
\newcommand{\np}{\texttt{numpy}}
\newcommand{\sci}{\texttt{scipy}}
\newcommand{\sy}{\texttt{sympy}}
\newcommand{\plt}{\texttt{matplotlib}}
\newcommand{\cpp}{\texttt{C++}}

\begin{document}

\journal{Computer Physics Communications}

\title{The \qs{} Python package: A physics toolbox for laser spectroscopy}

\author[1]{P. M\"uller\orcidlink{0000-0002-4050-1366}\corref{cor}}\ead{pmueller@physics.ucla.edu}\fnref{UCLA}
\author[1,2]{W. N\"ortersh\"auser\orcidlink{0000-0001-7432-3687}}

\cortext[cor]{Corresponding author}
\fntext[UCLA]{Present address: Department of Physics \& Astronomy, University of California Los Angeles, Los Angeles, California 90095, USA}

\affiliation[1]{
  organization={Institut f\"ur Kernphysik, Technische Universit\"at Darmstadt},
  postcode={64289},
  city={Darmstadt},
  country={Germany}}
  
\affiliation[2]{
  organization={Helmholtz Research Academy Hesse for FAIR, Campus Darmstadt},
  postcode={64289},
  city={Darmstadt},
  country={Germany}}

\begin{abstract}

The analysis of experimental results with Python often requires writing many code scripts which all need access to the same set of functions. In a common field of research, this set will be nearly the same for many users. The \qs{} Python package was developed to provide functions for physical formulas, simulations and data analysis routines widely used in laser spectroscopy and related fields. Most functions are compatible with \np{} arrays, enabling fast calculations with large samples of data. A multidimensional linear regression algorithm enables a King plot analyses over multiple atomic transitions. A modular framework for constructing lineshape models can be used to fit large sets of spectroscopy data. A simulation module within the package provides user-friendly methods to simulate the coherent time-evolution of atoms in electromagnetic fields without the need to explicitly derive a Hamiltonian.

\end{abstract}

\begin{keyword}
Laser spectroscopy \sep Python \sep Data analysis \sep Multidimensional linear regression \sep Maximum likelihood fitting \sep Master equation
\end{keyword}

\maketitle

\section{Introduction}
\label{sec:introduction}

Laser spectroscopy is in general concerned with the manipulation of the inner and outer degrees of freedom of atoms or ions using laser light. The subject of the interaction reaches from individual ions or atoms captured in a trap, over hot thermal ensembles in a neutral or charged plasma up to beams at high and even relativistic speed. The goal of the laser interaction can also be divers, \eg, transferring atoms into a specific (excited) state, generating fluorescence photons for detection, change the ions motional degree by momentum transfer (laser cooling), or to produce nuclear or atomic polarization or alignment.
The \qs{} package presented here is intended to provide a fundamental physics and data processing framework for experiments based on laser-atom interactions, where atoms are understood as electrons bound to a moving unresolved nucleus. While parts of \qs{} are made for general data processing and physics calculations, being based in laser spectroscopy experiments, the generic experiment benefiting the most from \qs{} produces data by reading out the response of atoms after manipulating them with lasers.
\noindent Ultimately, these are experiments whose means are the determination of high-precision transition frequencies for a wide variety of possible goals, which range from collinear laser spectroscopy (CLS) to probe atomic or nuclear structure theory \cite{Neugart.1981, Otten.1989, Blaum.2013, Campbell.2016, Ullmann.2017, Mueller.2020}, over searches for new physics \cite{Hudson.2011, Safronova.2018}, to the development of atomic clocks \cite{Bloom.2014, Ludlow.2015, Campbell.2017} or quantum computing with atom or ion traps \cite{Cirac.1995, Haeffner.2008, Friis.2018}. During the planning or the analysis phase, most of these experiments require simulations of the time evolution of atoms in laser fields, nonlinear fits to spectroscopy data, error propagation or simple calculations of laser frequencies, power, polarization as well as energies, velocities, Doppler, hyperfine-structure (hfs), or Zeeman shifts \etc{}\\
\noindent The goal of the \qs{} package is to make calculations of such observables and simulations easily accessible in any analysis script without the need to copy source code into every new project. While this is already useful for the most basic functions, it is even more practical for more extensive calculations, as long as they can still be defined in sufficiently general terms. In this sense, the package has been widely used and tested in simulating and analyzing laser spectroscopy results in collinear laser spectroscopy at COALA \cite{Koenig.2020}, COLLAPS/ISOLDE \cite{Neugart.2017}, and spectroscopy at storage rings \cite{Mohr.2024.arxiv}, but can be easily used for laser spectroscopy on thermal beams or in ion or atom traps as well.\\
\qs{} expands on the concept of assisting the planning and analysis of experiments by also providing general mathematical, statistical and optimization methods, building on the foundation of the \np{} and \sci{} packages. Additionally, object-oriented frameworks to create fit models in a modular way (\qmod{}) and to time-evolve coherent laser-atom interactions (\qsim{}) are implemented. These two frameworks and the derivation of a maximum-likelihood fit of a straight line to $n$-dimensional uncertain data points constitute the main body of this article. In the following section, a brief introduction into atomic hyperfine structure and fluorescence spectra is given, as these constitute the core of the implemented theory of light-matter interactions. In Sec.\,\ref{sec:structure}, a technical summary and an overview over the \qs{} package is given, including a short tutorial on how to get started.

\section{Theory}
\label{sec:theory}

In atomic physics, the required theory and calculations are often well understood and developed such that the user of a physics code library can quickly relate a name of a method to the theory and asses the underlying physical assumptions and approximations. Hence, for introductions to the general physical concepts and the standard approaches to laser spectroscopy and the underlying atomic physics of the \qs{} package, standard textbook resources can be consulted \cite{Drake.2023, Loudon.2000, Demtroeder.2008}. However, in high-precision experiments, also small effects such as hyperfine-induced mixing \cite{Johnson.1997} or quantum interference in the photon scattering rate \cite{Brown.2013, Beyer.2017} can significantly influence an experiment. Since these effects go beyond the standard frameworks, but are also within the scope of \qs{}, they shall be briefly introduced in the following paragraphs.\\
\noindent In the field of laser spectroscopy, usually transition frequencies between fine-structure states are addressed, \eg, for determining isotope shifts, or are specified in literature, while hyperfine-structure splittings, emerging in isotopes with nonzero nuclear spin, often need to be calculated. The hyperfine structure Hamiltonian, describing the higher-order multipole terms of the electromagnetic interaction between the electrons and the nucleus beyond the Coulomb interaction, can be written as \cite{Schwartz.1955, Drake.2023}
\begin{align}
\label{eq:hfs-hamiltonian}
H_\mathrm{HFS} &= \sum\limits_{k\geq 1}\bm{M}^{(k)}\cdot\bm{T}^{(k)} \equiv \sum\limits_{k\geq 1}\sum\limits_{\lambda=-k}^k (-1)^{\lambda} M_{-\lambda}^{(k)}\cdot T_\lambda^{(k)},
\end{align}
where $\bm{M}^{(k)}$ and $\bm{T}^{(k)}$ are irreducible tensor operators of rank and multipole order $k$, acting on the nucleus and the electrons, respectively. In the second expression, the scalar product is written out in the spherical basis. Using the Wigner-Eckhart theorem, this can be written as
\begin{align}
\label{eq:hfs-matrix}
h\nu^{F}_{(\gamma J)(\gamma^\prime J^\prime)}&\coloneqq\braket{\alpha I, \gamma J; Fm| H_\mathrm{HFS}|\alpha^\prime I^\prime, \gamma^\prime J^\prime; F^\prime m^\prime}\nonumber\\
& = \sum\limits_{k\geq 1}\delta_{FF^\prime}\delta_{mm^\prime}(-1)^{I^\prime + J + F}\begin{Bmatrix}I & J & F\\J^\prime & I^\prime & k\end{Bmatrix}\nonumber\\
&\qquad\times\braket{\alpha I||\bm{M}^{(k)}||\alpha^\prime I^\prime}\braket{\gamma J|| \bm{T}^{(k)}||\gamma^\prime J^\prime},\\
\braket{II|M_0^{(k)}|II} &\eqqcolon \begin{pmatrix}I & k & I\\I & 0 & -I\end{pmatrix}\braket{I||\bm{M}^{(k)}||I}\\
\braket{JJ|T_0^{(k)}|JJ} &\eqqcolon \begin{pmatrix}J & k & J\\J & 0 & -J\end{pmatrix}\braket{J||\bm{T}^{(k)}||J}
\end{align}
where $h$ is the Planck constant, $\delta_{FF^\prime}$, $\delta_{mm^\prime}$ are Kronecker-deltas, $(:::)$ and $\lbrace:::\rbrace$ are the Wigner-3j and -6j symbols and reduced matrix elements are denoted by $\braket{\cdot||\cdot||\cdot}$. The quantum numbers $I$, $I^\prime$ represent the nuclear spin, $J$, $J^\prime$ the total electron angular momentum, $F$ the combined total angular momentum and $\alpha$, $\gamma$ encapsulate additional quantum numbers such as particle spins. If the off-diagonal elements, linking different $I$ and $J$, vanish, this expression can be simplified to the textbook hyperfine structure formula \cite{Schwartz.1955, Drake.2023}
\begin{align}
\label{eq:hfs}
\nu^{F}_{(\gamma J)(\gamma J)} &= A\frac{K}{2} + B\frac{\frac{3}{4}K(K + 1) - I(I + 1)J(J + 1)}{2I(2I - 1)J(2J - 1)}\\[1ex]
&\hspace{-2em} + C\frac{\left[\splitdfrac{\frac{5}{4}K^3 + 5K^2 - 5I(I + 1)J(J + 1)}{+ K(I(I + 1) + J(J + 1) - 3I(I + 1)J(J + 1) + 3)}\right]}{I(I - 1)(2I - 1)J(J - 1)(2J - 1)}\nonumber\\[3ex]
K &\coloneqq F(F + 1) - I(I + 1) - J(J + 1)\\
A_k &\coloneqq \braket{II|M_0^{(k)}|II}\braket{JJ|T_0^{(k)}|JJ}\\
hA &\coloneqq \frac{A_1}{IJ},\qquad hB \coloneqq 4A_2,\qquad hC \coloneqq A_3,
\end{align}
\noindent
with the hyperfine structure constants $A$, $B$ and $C$ in frequency units for the magnetic dipole, electric quadrupole and magnetic octupole term, respectively. Off-diagonal elements with respect to $I$ are often neglected due to the large energy gaps between nuclear states, but will be of interest in very rare cases like $^{229}$Th, which has a particularly low-lying isomeric state that leads to hyperfine-induced nuclear-level mixing with dramatic consequences for the lifetime of the nuclear state \cite{Shabaev.2022}. Off-diagonal terms in $J$ are much more common and can give rise to hyperfine-induced fine-structure mixing, as for example in $^3$He \cite{Liao.1980} or $^{11}$B$^{3+}$ \cite{Mohr.2023}. If this is the case, the full Hamiltonian has to be diagonalized block-wise for each $F$ to extract individual fine-structure frequencies $\nu_J$, \ie{}, the matrices
\begin{align}
\label{eq:hfs-mixing}
N^F_{(\gamma J)(\gamma^\prime J^\prime)} = \nu_J\delta_{(\gamma J)(\gamma^\prime J^\prime)} + \nu^{F}_{(\gamma J)(\gamma^\prime J^\prime)}
\end{align}
need to be diagonalized. The \qs{} package enables fitting to spectra with mixed $J$ through numerical diagonalization. However, this requires additional input from theory or experiment. The nuclear moments, as well as the off-diagonal matrix elements $\braket{\gamma J|| \bm{T}^{(k)}||\gamma^\prime J^\prime}$ need to be provided. Nuclear magnetic dipole moments ($k=1$) can be found in \cite{Stone.2019}, electronic magnetic dipole matrix elements for He-like systems are listed, \eg{}, in \cite{Johnson.1997}. The implementation in \qs{} is currently limited to $k=1$, but the extension is straightforward and may follow in a future update.\\
Fluorescence spectra are generated from detected photons resonantly scattered from an atom. Resonance peaks appear where the frequency of the incident photon matches that of an atomic transition. If excited quantum states are energetically close, a fluorescence spectrum cannot be simply described as a sum of resonance peaks anymore, but the superposition of the excited states has to be considered. This is particularly relevant for unresolved hyperfine structure spectra \cite{Blaum.2013}, but can also become relevant in high-precision measurements were peaks are hundreds of linewidths separated \cite{Beyer.2017}. This ``quantum interference'' (QI) effect depends on the solid angle of detection of the emitted fluorescence light and leads in general to asymmetric lineshapes, which leads to shifts in the determined transition frequencies if simply fitted with a sum-of-peaks model. In \qs{}, the QI scattering rate derived in \cite{Brown.2013} is implemented as a fit model of the form
\begin{align}
\label{eq:scattering rate pert}
\frac{\mathrm{d}\Gamma}{\mathrm{d}\Omega}(\bm{k}_\mathrm{sc}) &\coloneqq \frac{1}{4\pi}\frac{I}{I_0}\left(\frac{\Gamma}{2}\right)^3\Bigg[\sum_{F^\prime}\frac{f(\bm{q}, \bm{k}_\mathrm{sc}, F, F^\prime)}{\Delta_{FF^\prime}^2 + (\Gamma/2)^2} \nonumber\\[1ex]
&\quad + \sum_{\substack{F^\prime, F^{\prime\prime}\\ F^\prime \neq F^{\prime\prime}}}\frac{g(\bm{q}, \bm{k}_\mathrm{sc}, F, F^\prime, F^{\prime\prime})}{(\Delta_{FF^\prime} + \frac{i}{2}\Gamma)(\Delta_{FF^{\prime\prime}} - \frac{i}{2}\Gamma)}\Bigg]
\end{align}
\begin{align}
f(\bm{q}, \bm{k}_\mathrm{sc}, F, F^\prime) &= A^{F^\prime}_{F} + B^{F^\prime}_{F}p(\bm{q}, \bm{k}_\mathrm{sc})\\
g(\bm{q}, \bm{k}_\mathrm{sc}, F, F^\prime, F^{\prime\prime}) &= C^{F^\prime F^{\prime\prime}}_{F}p(\bm{q}, \bm{k}_\mathrm{sc})\\
I_0 &= \frac{\hbar\Gamma\omega^3}{12\pi c^2},
\end{align}
where $p(\bm{k}_\mathrm{sc}, \bm{q})$ can be used as a fit parameter, takes values in $[-0.5, 1]$ and depends on the polarization of incident and the directions of detectable scattered photons. $I_0$ is the saturation intensity, $\hbar$ the reduced Planck constant, $c$ the vacuum speed of light, $\Delta_{FF^\prime} \coloneqq \omega_{F^\prime} - \omega_F$, $\omega\coloneqq 2\pi\nu$ and $\Gamma$ is the natural linewidth of the electronic transition. The coefficients $A^{F^\prime}_{F}$, $B^{F^\prime}_{F}$ and $C^{F^\prime F^{\prime\prime}}_{F}$ only depend on angular momentum algebra and are derived in \cite{Brown.2013}. For simulations, as part of \qsim{}, the scattering rate was extended to address individual magnetic substates, allowing the use of a magnetic field. If the perturbative scattering rate is not sufficient, the \qsim{} module can be used to coherently evolve an atom in a laser field using a master-equation approach. The resulting density matrix can be used to calculate the expectation value of a detection operator that gives the full non-perturbative solution and converges to the perturbative result for low laser powers and closed transitions in the equilibrium state. This is described in more detail in Sec.\,\ref{sec:interaction of light and matter}.

\section{Structure}
\label{sec:structure}

The \qs{} package provides functions for physical calculations and simulations such as the aforementioned examples in a user-friendly and efficient way. Whenever useful, functions are compatible with the \texttt{ndarray} class of \np{} \cite{numpy}. This enables fast calculations with large samples of data. In addition to \np{}, a small number of established scientific Python packages are required with \sci{} \cite{scipy}, \sy{} \cite{sympy} and \plt{} \cite{matplotlib}. Parts of \qs{} are written in \cpp{} and are currently only compiled for Windows systems in a Dynamic Link Library (DLL). A detailed overview of \qs{} with installation instructions, tutorials, the API documentation and links to the Python Package Index (PyPI), the GitHub repository and this publication can be found on the official website \cite{qspec}. Supplemental details are also given in \cite{Mueller.2024.Dissertation}.\\
\noindent The content of the \qs{} package is distributed over seven largely independent modules/categories. These are named \texttt{analyze}, \texttt{algebra}, \texttt{models}, \texttt{physics}, \texttt{simulate}, \texttt{stats}, and \texttt{tools}. However, all but the \texttt{models} and the \texttt{simulate} modules can be accessed by simply importing \qs{}. This is demonstrated in the following short code example, often required in collinear laser spectroscopy. In this technique, a fast beam of ions or atoms is superimposed with a copropagating or counterpropagating laser beam \cite{Kaufman.1976} and a resonance curve is usually recorded by Doppler-tuning, \ie{}, changing the ion velocity with a small potential change $\Delta U$. The laboratory laser frequency required to excite the $24516.65$\,cm$^{-1}$ transition in $^{88}$Sr$^+$ ions, accelerated to $20$\,keV, with a counterpropagating laser beam is calculated in the following example together with the change in the rest-frame frequency of the ion for a potential variation of $1$\,V. Note that, all of the scalar values in this example could be replaced by \np{} arrays.
\begin{python}
import qspec as qs
# This imports the analyze, algebra,
# physics, stats, and tools modules

q = 1  # (e), Ion charge state
m = 87.905612253 - q * qs.me_u  # (u)
# Mass of 88Sr+

U = 20000  # (V), Acceleration voltage

# Resonance frequency from NIST database
f0 = qs.inv_cm_to_freq(24516.65)  # (MHz)
# >>> 734990676.5 MHz

# Relativistic velocity of 88Sr+
v = qs.v_el(U, q, m)  # (m/s)
# >>> 209533.6 m/s

# The anti-collinear lab. frequency
f_laser = qs.doppler(
    f0, v, qs.pi, return_frame='lab')  # (MHz)
# >>> 734477149.8 MHz

# The differential Doppler shift
df_atom = qs.doppler_el_d1(
    f_laser, qs.pi, U, q, m)  # (MHz / V)
# >>> +12.84 MHz / V
\end{python}
Functions are generally named according to their returned observable and required input parameters. Similar to this simple example, all functions and classes can be accessed, including the $n$-dimensional linear fit algorithm described in Sec.\,\ref{sec:multidimensional linear regression}. Elaborate introductions to the two separate frameworks \qmod{} and \qsim{} are given in Sec.\,\ref{sec:modular fit models} and \ref{sec:interaction of light and matter}, respectively. The following paragraphs give a short overview over the seven modules/categories.

\subsection{algebra}
\label{sec:algebra}

The \texttt{algebra} module is a collection of functions related to angular momentum coupling such as Wigner-$j$ symbols, reduced dipole matrix elements and relative transition strengths for electronic transitions. This is the only module that uses the \sy{} package to optionally return algebraic expressions of coupling strengths.

\subsection{analyze}
\label{sec:analyze}

The \texttt{analyze} module contains general purpose routines for (non)linear data fitting as well as a convenience class for multi-dimensional King plots \cite{King.1984}. Wrappers for the \texttt{scipy.optimize.curve\_fit} and the \texttt{scipy.odr.odr} routine, which use least square optimization and orthogonal distance regression, respectively \cite{scipy}, extend their functionality and standardize their syntax. For fitting of a straight line, as needed in King-plots, the York algorithm described in \cite{York.2004} and the Monte-Carlo (MC) method described in the supplementary material of \cite{Gebert.2015} are implemented. In both cases, also a generalized version is implemented for fitting a straight line to data points in multiple dimensions, whose uncertainties are determined by independent multivariate normal distributions. The multidimensional fit algorithms are described in Sec.\,\ref{sec:multidimensional linear regression}.

\subsection{models}
\label{sec:models}

The \texttt{models} module contains classes and helper functions to create lineshape models of fluorescence spectra or more general fit models for nonlinear curve-fitting. This module is similar to the existing \stt{} Python package \cite{satlas, satlas2}, which is widely used by laser spectroscopy groups at KU Leuven, CERN/ISOLDE and JYVL. In \texttt{models}, additional features such as numerical convolution with commonly used probability density functions, quantum interference (QI) effects \cite{Brown.2013} or hyperfine-induced mixing \cite{Johnson.1997} are available. The \texttt{models} module is described in detail in Sec.\,\ref{sec:modular fit models}

\subsection{physics}
\label{sec:physics}

The \texttt{physics} module provides physical relations, unit conversions and observables surrounding (collinear) laser spectroscopy such as particle energy and velocity, Doppler, photon recoil, hyperfine structure and Zeeman shifts, refractive indexes, \etc{}

\subsection{simulate}
\label{sec:simulate}

The \texttt{simulate} module offers an intuitive framework for simulations of laser-atom interactions and is comparable to other packages that simulate the time evolution of quantum systems such as \qu{} \cite{Johansson.2012, Johansson.2013} or \texttt{PyLCP} \cite{Eckel.2020}. Currently, solvers for the Lindblad master, rate and Schr\"odinger equation as well as a MC solver for the Lindblad master equation are included. The differential photon scattering rate, which also includes quantum interference effects, can be calculated from the density matrix or the state vector. Additionally, a perturbative scattering rate can be calculated, as described in \cite{Brown.2013}. The \qsim{} module is described in more detail in Sec.\,\ref{sec:interaction of light and matter}.

\subsection{stats}
\label{sec:stats}
  
The \texttt{stats} module contains routines for the statistical analysis of data. Besides basic statistical measures such as the (weighted) average or median, the module also provides a MC error propagator. In combination with the nonlinear curve fitters of the \textit{analyze} module, this can be used to calculate uncertainty bands for the fitted function while taking the full covariance matrix into account.

\subsection{tools}
\label{sec:tools}

The \texttt{tools} module provides methods for data processing and conversion as well as general mathematical functions. Some examples are the conversion between lists and dictionaries, merging intervals, rounding to $n$ significant decimal places, transforming vectors or printing colored text to the console \etc{}

\section{Multidimensional linear regression}
\label{sec:multidimensional linear regression}
\begin{figure}[!b]
\centering
\includegraphics[width=\linewidth]{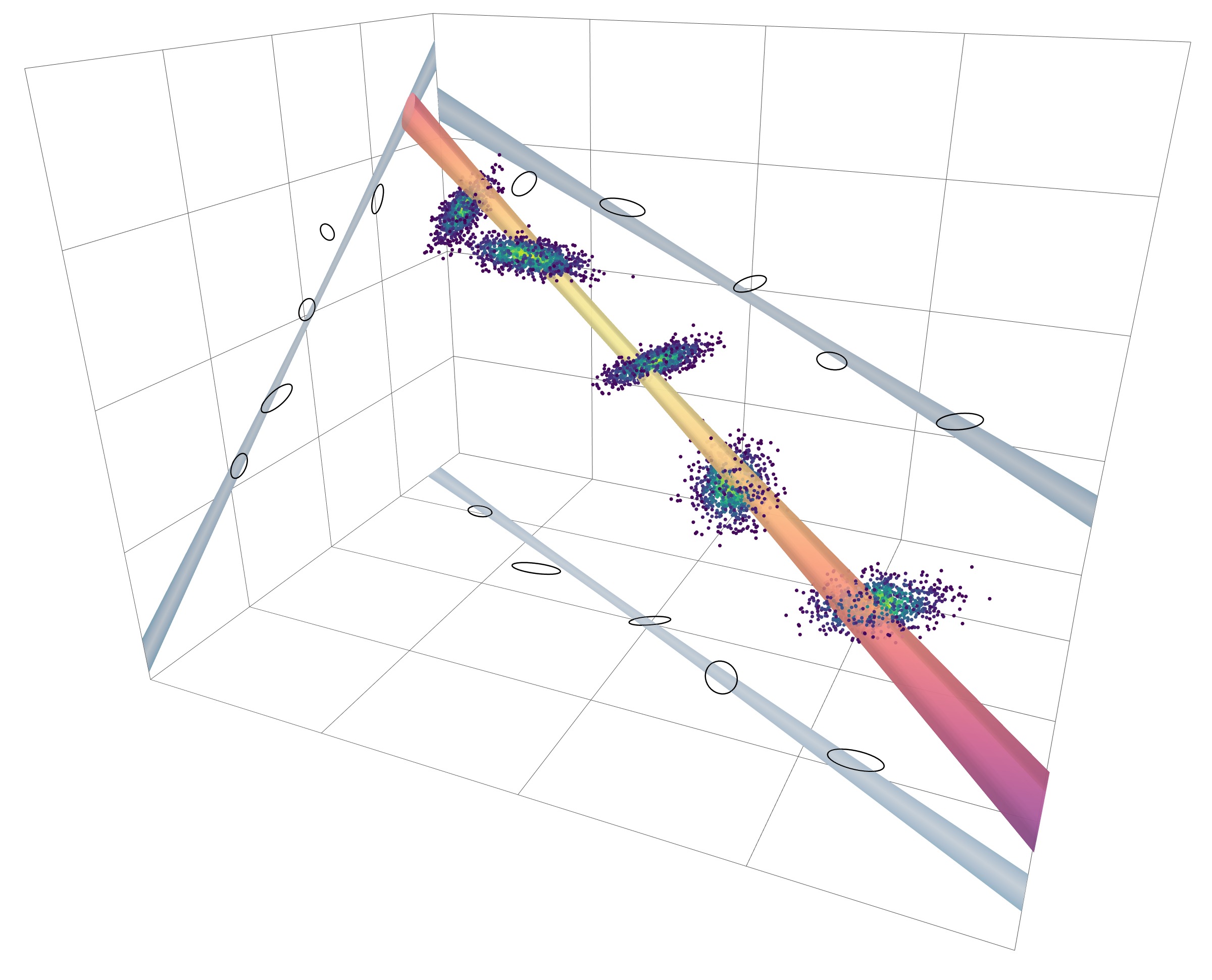}
\caption{Fit of a straight line to $5$ data points in $3$ dimensions. The scatter plots indicate the uncertainty regions of the data points. In the Monte-Carlo (MC) method, similar random samples are generated to construct a distribution of sample straight lines. The $1\sigma$ uncertainty region of the straight line fitted with the analytical algorithm is shown as a yellow-purple colormap. The projections of the $1\sigma$ uncertainty regions of the data points and the fitted straight line are shown on the three orthogonal planes.\label{fig:straight}}
\end{figure}
The \qs{} package provides an analytical and a MC method for fitting a straight line to data points in $k\in\mathbb{N}$ dimensions. For any set of data points $(\vec{X}_i)_{i\in\mathcal{I}}$, with the index set $\mathcal{I}$, the algorithms described here assume that the real value of the random vector $\vec{X}_i$ is distributed according to an independent multivariate normal distribution $\mathcal{N}(\vec{\mu}_i, \Sigma_i)$, with the mean vector $\vec{\mu}_i\in\mathbb{R}^k$ and the positive-semidefinite covariance matrix $\Sigma_i\in\mathbb{R}^{k\times k}$. Here, we distinguish between an uncertain data point $\vec{X}_i$, which may be the result of a measurement, and the actual or real but unknown value $\vec{x}_i$. Figure \ref{fig:straight} depicts the optimization problem in $3$ dimensions with the resulting straight line fit.\\
The MC approach generates sample vectors $\vec{x}_i$ according to the distributions $\mathcal{N}(\vec{\mu}_i, \Sigma_i)$ under the condition that they are aligned on a straight line. Accordingly, the result of the MC method are samples of straight lines following the probability density function imposed by the uncertain data points and the linear relation. The algorithm is described in detail in the supplemental material of \cite{Gebert.2015} for the $3$-dimensional case. However, the extension of this algorithm to $k$ dimensions is straightforward and is, thus, not described here.\\
The analytical approach is a maximum likelihood fit given the above described assumptions. Its result corresponds to the most probable sample straight line generated by the MC method. Note that in general this straight line is not given by the mean values of the slopes and $y$-intercepts of the samples. The algorithm described by York\etal{}~\cite{York.2004} yields the analytical solution for $2$ dimensions. Here, the extension to $k$ dimensions is not obvious. Hence, in the following, the solution developed for \qs{} is described.\\
Let a straight line in $k$ dimensions be parameterized by
\begin{align}
f[\vec{a}, \vec{b}]: \mathbb{R}\rightarrow \mathbb{R}^k,\quad t \mapsto \vec{a} + t\vec{b}.
\end{align}
%
%where $\mathcal{F} \coloneqq \lbrace f[\vec{a}, \vec{b}](t)\ |\ t\in\mathbb{R}\rbrace$ is the set of all points lying on the straight line.
Then we want to find the offset vector $\vec{a} \in \mathbb{R}^k$ and the directional vector $\vec{b} \in \mathbb{R}^k$ that define the most likely straight line through the set of uncertain data points $(\vec{X}_i)_{i\in\mathcal{I}}$, which generate the independent multivariate normal distributions $\mathcal{N}(\vec{\mu}_i, \Sigma_i)_{i\in \mathcal{I}}$. The probability density of $\vec{X}_i$ can be written as
\begin{align}
\label{eq:multivariate normal}
p_i(\vec{x}_i) &= \frac{1}{\sqrt{(2\pi)^k|\Sigma_i|}}\exp\left[ -\frac{1}{2}(\vec{x}_i - \vec{\mu}_i)^\mathrm{T}\Sigma_i^{-1}(\vec{x}_i - \vec{\mu}_i)\right]
\end{align}
where $^\mathrm{T}$ indicates transposition and $|\Sigma_i|$ and $\Sigma_i^{-1}$ are the determinant and the inverse of the covariance matrix, respectively. We can now define the likelihood of observing a sample $\vec{x}_i$ if it is constrained to the straight line $f[\vec{a}, \vec{b}]$
\begin{align}
&L_i(\vec{a}, \vec{b}\,|\,t)\coloneqq p_i(f[\vec{a}, \vec{b}](t))\nonumber\\
&\quad = \frac{1}{\sqrt{(2\pi)^k|\Sigma_i|}}\exp\left[ -\frac{1}{2}\left(\vec{a}_i^\mathrm{T}\Sigma_i^{-1}\vec{a}_i - \frac{t_i^2}{\sigma_i^2} - \frac{(t - t_i)^2}{\sigma_i^2}\right)\right]
\label{eq:likelihood-1}
\end{align}
with the definitions
\begin{align}
\vec{a}_i \coloneqq \vec{a} - \vec{\mu}_i,\quad \sigma_i \coloneqq \frac{1}{\sqrt{\vec{b}^\mathrm{T}\Sigma_i^{-1}\vec{b}}},\quad t_i \coloneqq -\sigma_i^2 \vec{a}_i^\mathrm{T}\Sigma_i^{-1}\vec{b}.
\end{align}
The most likely straight line can be found by maximizing this likelihood function with respect to $\vec{a}$ and $\vec{b}$. However, for any given straight line $f[\vec{a}, \vec{b}]$, the real value of $t$ that generates $\vec{x}_i$ is unknown. Using
\begin{align}
&\max\limits_{\vec{a}, \vec{b}} L_i(\vec{a}, \vec{b}\,|\,t) \leq \max\limits_{\vec{a}, \vec{b}} L_i(\vec{a}, \vec{b}\,|\,t = t_i)\nonumber\\
&\quad =\max\limits_{\vec{a}, \vec{b}}\frac{1}{\sqrt{(2\pi)^k|\Sigma_i|}}\exp\left[ -\frac{1}{2}\left(\vec{a}_i^\mathrm{T}\Sigma_i^{-1}\vec{a}_i - \frac{t_i^2}{\sigma_i^2}\right)\right],
\end{align}
and recalling that all $(\vec{X}_i)_{i\in\mathcal{I}}$ are independent by definition, the likelihood function that needs to be maximized to find the desired straight line through all data points is given by
\begin{align}
\label{eq:likelihood-2}
&L(\vec{a}, \vec{b}) \coloneqq \prod\limits_{i\in\mathcal{I}} L_i(\vec{a}, \vec{b}\,|\,t = t_i).
\end{align}
Analogously to standard least-square optimization algorithms, this is most efficiently done by minimizing
\begin{align}
\mathcal{L} &\coloneqq -\log(L(\vec{a}, \vec{b}))\nonumber\\
&= \frac{1}{2}\sum\limits_{i\in\mathcal{I}}\left[\log((2\pi)^k|\Sigma_i|) + \vec{a}_i^\mathrm{T}\Sigma_i^{-1}\vec{a}_i - \frac{t_i^2}{\sigma_i^2}\right],
\label{eq:log-likelihood}
\end{align}
where the terms $\log((2\pi)^k|\Sigma_i|)$ can be omitted as they are independent of $\vec{a}$ and $\vec{b}$. In \qs{}, the minimization is performed using the \texttt{scipy.optimize.minimize} function with the Newton conjugate gradient (Newton-CG) method \cite{scipy, Nocedal.2006}. For the fastest possible convergence, both the gradient and the Hessian of Eq.\,\eqref{eq:log-likelihood} are computed analytically, see \ref{sec:gradient and hessian of the negated log-likelihood function}. The covariance matrix of the resulting $\vec{a}$ and $\vec{b}$ are calculated by numerically inverting the Hessian at the solution with \texttt{numpy.linalg.inv}. For two dimensions, the algorithm described here reproduces the fit performed with the York algorithm to machine precision. The estimated uncertainties deviate slightly in the third significant digit.\\
The \qs{} package also contains a class to set up King plots where usually the data vectors, consisting of isotope shifts or nuclear charge radii, are multiplied by mass factors to yield modified data vectors according to
\begin{align}
\label{eq:x-mod}
\vec{x}_\mathrm{mod}^{\,AA^\prime} = \mu^{AA^\prime}\vec{x}^{\,AA^\prime} = \frac{(M^A + m_\mathrm{e})(M^{A^\prime} + m_\mathrm{e})}{M^{A^\prime} - M^A}\,\vec{x}^{\,AA^\prime},
\end{align}
where $M^A$ is the mass of the nucleus of the isotope with mass number $A$ and $m_\mathrm{e}$ is the mass of the electron. The modified data vectors $\vec{x}_\mathrm{mod}$ are then aligned on a straight line. In this case, the covariance matrices $\Sigma_i$ are calculated from the masses and data vectors $\vec{x}^{\,AA^\prime}$. This yields notable positive correlations between the vector components if the uncertainties of the masses are comparable or larger than those of the data vectors. Correlations between different data vectors are neglected. Improved or new data is gained from using the fitted straight line to calculate missing vector components $j$ from one known component $l$
\begin{align}
x_\mathrm{mod}^j = a^j + \frac{b^j}{b^l}\left(x_\mathrm{mod}^l - a^l\right).
\end{align}
The improved unmodified data and their covariance matrix is determined by inverting Eq.\,\eqref{eq:x-mod} and using Gaussian error propagation. A code example of a simple 2d King plot with Ca$^+$ isotopes is given in \mbox{Code Listing \ref{code:king-plot}}, with data from \cite{Mueller.2020, GarciaRuiz.2016, Wang.2021}.

\section{Modular fit models}
\label{sec:modular fit models}

In \qmod{}, a modular system for constructing fit models, with a focus on lineshape models, is implemented. This module can be compared to the \stt{} package, which is used for fitting lineshape models to low- and high-statistics data \cite{satlas, satlas2}. In the backend, \qmod{} extends the \texttt{scipy.optimize.curve\_fit} method for fitting \cite{scipy} while \stt{} uses the \texttt{lmfit} package \cite{Newville.2024}. Both packages provide a modular system for constructing nonlinear functions, routines for fitting data with low statistics and options to alter the influence of individual parameters. While \qmod{} implements models and features that are currently not available in \stt{}, we also acknowledge that \stt{} provides additional fit options such as Poisson-statistic maximum-likelihood fitting or Bayesian inference that are currently not available in \qmod{}. A complete up-to-date feature comparison is given in Tab.\,\ref{tab:models-features}.
\begin{table}[t]
\centering
\caption{\label{tab:models-features} List of features implemented in \stt{} or \qmod{}. Please note that this feature list is only a momentary capture, as both packages are in active development, but it serves as a useful overview. Here $(x, y)$ is a data point to be fitted, $\Delta y$ is the uncertainty of the $y$ value, $f$ is the fit model, $\lbrace p\rbrace$ is the set of fit parameters and $\rho$ returns the uncertainty $\Delta y$ of a data point at runtime.}
\resizebox{\columnwidth}{!}{%
\begin{tabular}{l c c}
Feature & \stt{} & \qmod{} \\[0.5ex]
\hline \\[-1.5ex]
Modular fit models & \checkmark & \checkmark\\
Quantum interference & --- & \checkmark\\
Hyperfine mixing & --- & \checkmark\\
Convoluted lineshapes & --- & \checkmark\\
Shared parameters & \checkmark & \checkmark\\
Parameters as functions & \checkmark & \checkmark\\
Priors & \checkmark & \checkmark\\[0.5ex]
\hline \\[-1.5ex]
Gaussian statistics & \checkmark & \checkmark\\
Poisson statistic & \checkmark & --- \\
$\Delta y$ as a function & $\rho(f(x, \lbrace p\rbrace))$ & $\rho(x, y, f(x, \lbrace p\rbrace), \lbrace p\rbrace)$\\
Bayesian inference & \checkmark & --- \\
\end{tabular}}
\end{table}

\noindent There are several reasons that have lead to the development of \qmod{}. At the beginning of its development, only the old version of \st{} was available \cite{satlas}. Therefore, a large performance boost was expected in \qmod{}, as evidenced by the now available faster new version \stt{} \cite{satlas2}. In \qmod{}, new features and physical models are implemented that were/are not available in \stt{}, such as consideration of quantum interference effects or convolutions of lineshape models. The dependencies of \qmod{} are kept at a minimum of only \np{} and \sci{}. Moreover, \qmod{} is now used as the backend of an updated version of the \texttt{Tilda.PolliFit} framework, which provides a graphical user interface and is already used for the analyses of online data \cite{Kaufmann.2019.Dissertation, MalbrunotEttenauer.2022, Mueller.2024}. In the following, the structure of \qmod{} is described and a performance comparison to \stt{} is given.\\
The basis for the modular system is the abstract \texttt{Model} class, from which all models inherit, see the class diagram in Fig.\,\ref{fig:models-uml}. Composite models are created by passing a model to the constructor of another model. A basic example could be the creation of a hyperfine-structure model for a transition from a lower state $\ket{J=\nicefrac{1}{2}}$ to an upper state $\ket{J^\prime=\nicefrac{3}{2}}$ in a system with nuclear spin $I=\nicefrac{1}{2}$
\begin{align}
\label{eq:fit model}
f(x) &= p_0\sum\limits_{F\rightarrow F^\prime} a(F, F^\prime) \mathrm{Voigt}(x - \bar{x}, \Gamma, \sigma) + y_0\\
\bar{x} &= x_0 + \frac{A^\prime}{2}[F^\prime(F^\prime + 1) - 9/2] - \frac{A}{2}[F(F + 1) - 3/2]\nonumber,
\end{align}
where $\Gamma$, $\sigma$, $A$, $A^\prime$, $a(F, F^\prime)$, $x_0$, $p_0$ and $y_0$ are parameters. This would correspond to the following lines of code.

\begin{python}
from qspec.models import \
    Voigt, Hyperfine, NPeak, Offset
    
I, J_l, J_u = 0.5, 0.5, 1.5  # = I, J, J'
hfs = Hyperfine(Voigt(), I, J_l, J_u)
f = Offset(NPeak(hfs, n_peaks=1))
\end{python}

\noindent Here, \texttt{Voigt} creates the peak shape model with parameters $\Gamma$ (Lorentzian FWHM) and $\sigma$ (Gaussian standard deviation) and \texttt{Hyperfine} introduces the sum in Eq.\,\eqref{eq:fit model}, the hyperfine-structure constants $A$, $A^\prime$ and the peak intensities $a(F, F^\prime)$. The \texttt{NPeak} model creates a sum over, in this case, one instance of the passed hyperfine-structure model at the position $x_0$ with intensity $p_0$. The Offset model creates the offset parameter $y_0$.\\
The lineshape model defined in Eq.\,\eqref{eq:fit model} can be substantially modified. The Voigt profile can be replaced by other peak shapes. The chosen peak shape can also be numerically convolved with a second peak shape of choice. The multipole expansion of the hyperfine structure is implemented up to magnetic octupole terms. Quantum interference (QI) effects can be considered by replacing \texttt{Hyperfine} with \texttt{HyperfineQI}, see Eq.\,\eqref{eq:scattering rate pert}. This creates an extra parameter that encapsulates the detector geometry determining the QI effect. Hyperfine-induced mixing is considered in \texttt{HyperfineMixed}. The \texttt{NPeak} model can be used to create additional copies of the passed submodel with individual positions and intensities. \texttt{Offset} can divide the $x$-axis into intervals with individual offset functions, which can be any polynomial. All parameters of a model can be free, fixed, functions of other parameters, bounded to an interval or statistically constraint by a reference value with uncertainty, also known as a \textit{prior} \cite{satlas2, Jaynes.1968}. Additionally, simultaneous fits of multiple data sets with shared parameters are possible. In a future update, \qmod{} can be combined with the \qsim{} module, described in Sec.\,\ref{sec:interaction of light and matter}, to enable fitting of non-perturbative scattering rates.
\begin{figure}[!t]
\centering
\includegraphics[width=\linewidth]{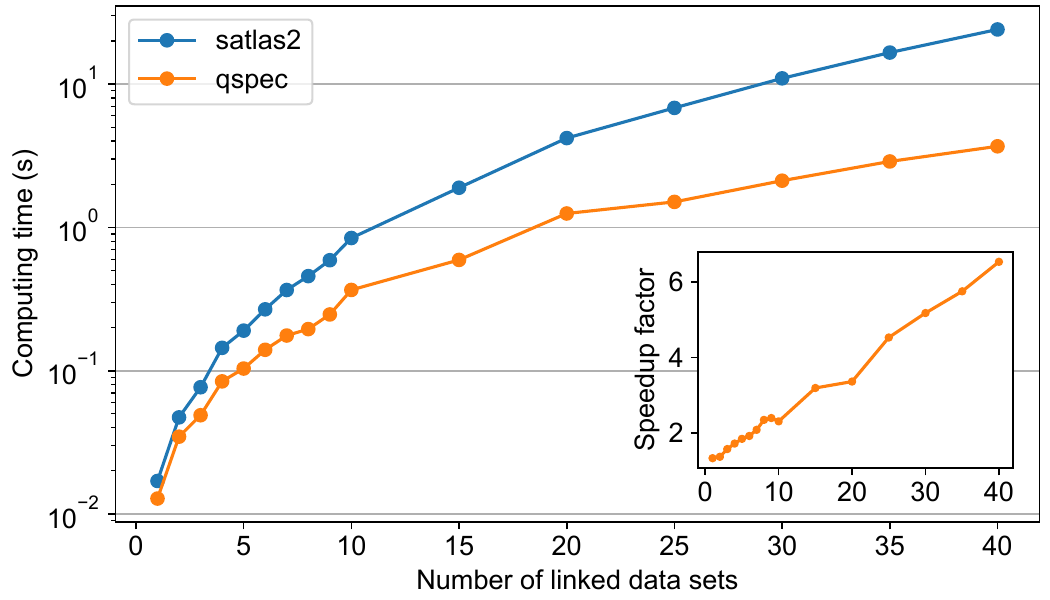}
\caption{Speed comparison of \qmod{} with \stt{} using an adaption of the benchmark test described in the documentation of \stt{}. The computing time increases polynomially with the number of linked data sets. The ratio of the computing times of \stt{} and \qmod{} grows linearly. The benchmark test was performed with \stt{} version 0.2.7 and \qs{} version 0.3.5 with an AMD Ryzen\texttrademark{} 7 7800X3D 8-Core Processor on Windows 11.\label{fig:benchmark}}
\end{figure}
\noindent The computing time for a fit with shared parameters increases polynomially with the number of linked data sets. Therefore, the fit models have to be calculated efficiently. To test the speed of \qmod{}, the benchmark test described in the documentation of \stt{} \cite{satlas2} was adapted, which compares \stt{} to \st{} for fits to different numbers of data sets using shared parameters. There, a speedup factor of two to three orders of magnitude was found, depending on the number of data sets \cite{satlas2}. The benchmark test is a fit of Voigt profiles to the hyperfine structure spectrum of a $\ket{J=\nicefrac{1}{2}}\rightarrow\ket{J^\prime=\nicefrac{3}{2}}$ transition with low Poisson statistics in a system with nuclear spin $\nicefrac{7}{2}$. The hyperfine-structure parameters $A$, $A^\prime$ and $B^\prime$ as well as the parameters $x_0$, $\Gamma$ and $\sigma$ were shared across data sets, compare Eq.\,\eqref{eq:fit model}. The magnetic octupole constant $C^\prime$ was fixed and set to zero. The low Poisson statistics were accounted for by a custom function
\begin{align}
\rho(x, y, f(x, \lbrace p\rbrace), \lbrace p\rbrace) = \sqrt{f(x, \lbrace p\rbrace)},
\end{align}
which computes the uncertainty of the data during runtime, see Tab.\,\ref{tab:models-features}. Figure \ref{fig:benchmark} shows a comparison of the computing times of \qmod{} and \stt{}. A linearly growing speedup factor of $1.2$ to $7$ was found when using \qmod{} for simultaneously fitting $1$ to $40$ randomly generated data sets, respectively. The fit results agree well and only deviate in the third significant digit of the uncertainties. Please note that these computing speed results are only valid for this specific test configuration, as the speedup factor can depend on many parameters such as the used fit algorithm, the number of fit parameters, the number of linked parameters or the used test machine. Moreover, \stt{} provides additional statistical measures for the fit results whose computing times we could not test. However, these results show that the basic framework of \qmod{} meets the required performance standards and additional features, currently not implemented in \qmod{}, such as Bayesian inference or maximum-likelihood fitting, can now be added to the robust basis. These will call for a separate benchmark test.

\section{Interaction of light and matter}
\label{sec:interaction of light and matter}

The lineshape of fluorescence spectra cannot always be described by an analytical model, especially when atomic states are optically pumped. In such cases it is desirable to simulate the evolution of the atomic state population in the laser field to understand the origin of the experimental lineshape and to investigate systematic frequency shifts when using a simple lineshape model. Moreover, the simulations can help to find the optimal experimental parameters in experiments that rely on optical pumping effects. All experiments that involve light-atom interactions can potentially benefit from such simulations. Some examples are saturation spectroscopy, Ramsey interferometry or quantum computing with \mbox{atom/ion traps}.\\
Python packages to simulate the interaction of light and matter already exist. However, they are often designed for a certain subspace of atomic systems and applications or require the user to derive the mathematical objects such as Hamiltonians or observable operators. For example, the \qu{} package is a powerful library of solvers for open quantum systems which takes the Hamiltonian of the system as an input \cite{Johansson.2012, Johansson.2013}. For the simulation of laser spectroscopy experiments, where multiple isotopes with different hyperfine structures are of interest, a lot of work is required to set up all Hamiltonians. The \texttt{PyLCP} package facilitates this process, provides classes to set up lasers and external static electromagnetic fields and also allows to simulate the motional degrees of freedom, \eg{}, to simulate laser cooling \cite{Eckel.2020}. While this package has a large range of applications, it still requires the user to construct a Hamiltonian from individual energy contributions and it currently does not support passing \np{} arrays of velocities or laser detunings to simulate large samples. As a consequence, iterations over the samples and multi-threading for better performance have to be implemented by the user.\\
The \texttt{qspec.simulate} module provides an intuitive class system, which is comparable to drawing an atomic level scheme, to define laser-atom interactions. A class diagram of \qsim{} is depicted in Fig.\,\ref{fig:simulate-uml}. Time-independent Hamiltonians in the interaction picture are generated automatically from the specified atom, lasers and environment. Initial populations of atomic states, initial velocities and laser detunings can be passed as \np{} arrays. The backend utilizes multi-threading and is written in \cpp{} for the best performance and long-term support. Solvers for the rate equations, the Schr\"odinger equation and the master equation are available. Additionally, a Monte-Carlo (MC) master equation solver can be used to also simulate the recoil motion of the atom in the laser field. In \texttt{qspec.simulate}, an atom is defined as a list of states $\ket{i}\coloneqq\ket{\gamma,S,L,J,I,F,m_F}$ with eigen energies $h\nu_i$ that are connected through spontaneous decay rates $\Gamma_{\!ij}$. The eigen frequencies $\nu_i$ are either specified directly or calculated from Eq.\,\eqref{eq:hfs} and the center-of-gravity frequency $\nu_0$ of the fine-structure states $\ket{\gamma,S,L,J}$.\\
Lasers are approximated as classical monochromatic plane waves, for which the user specifies the frequency $\tilde{\nu}_k$, intensity $I_k$ and complex polarization vector $\vec{q}^{\,k}$. The atom and the lasers are combined into an \texttt{Interaction} object which generates the differential equations to be solved. For a complete list of definitions for the mathematical objects used in \qsim{}, see \ref{sec:definitions for qspec.simulate}.

\subsection{Hamiltonian}
\label{sec:hamiltonian}

\begin{figure}[!t]
\centering
\includegraphics[width=\linewidth]{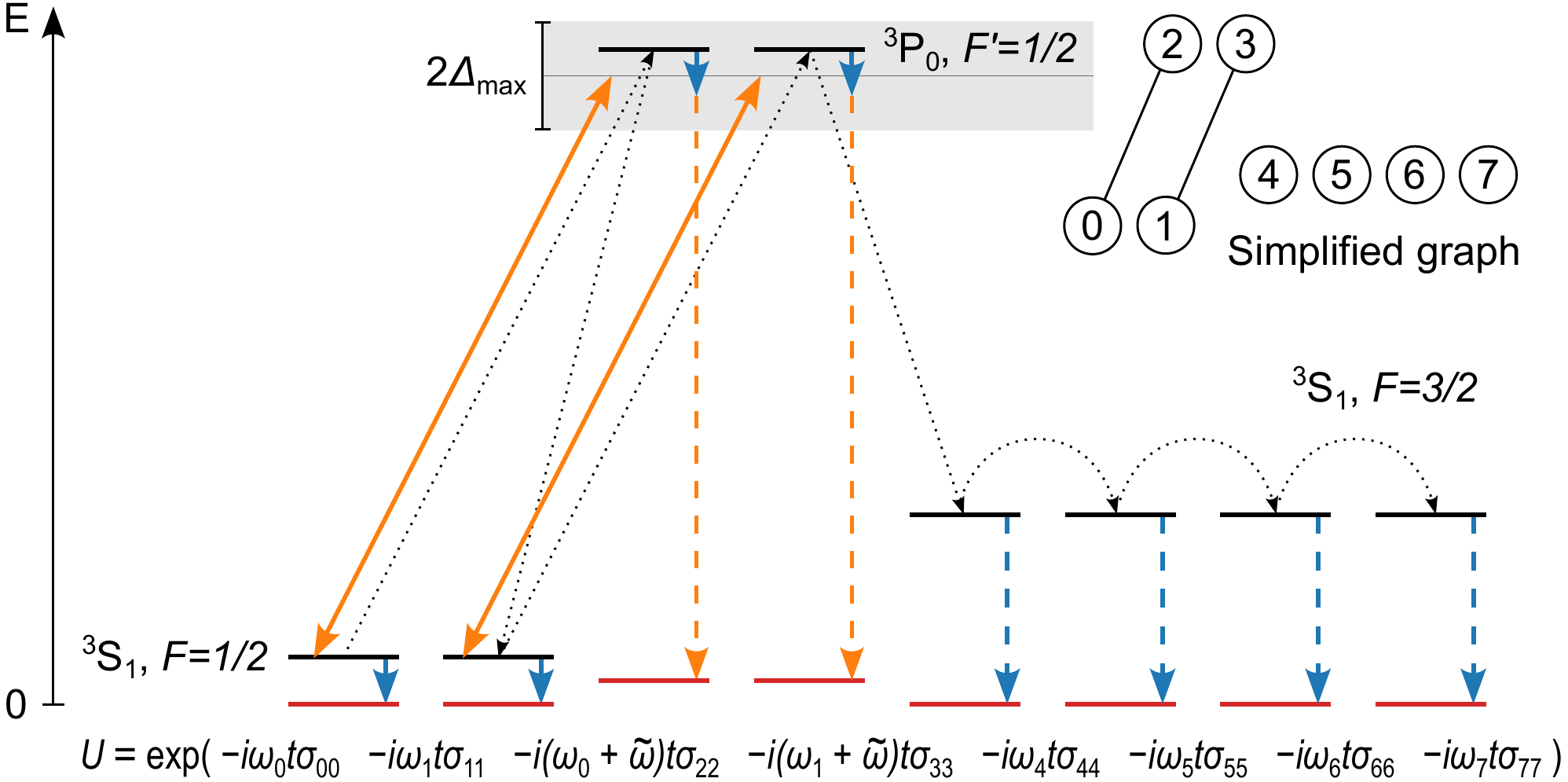}
\caption{Sketch of the algorithm used to transform the Hamiltonian into the interaction picture for a laser-driven $^3$S$_1\rightarrow\,^3$P$_0$ transition in a nuclear spin-$\nicefrac{1}{2}$ system. The solid black and red lines depict the energies of the atomic states (diagonal elements of the Hamiltonian) before and after the transformation, respectively. The solid orange arrows depict the lasers, which are slightly detuned from resonance. The dashed arrows indicate the transformations applied to each state. The transformations applied to the $^3$S$_1(F=\nicefrac{1}{2})$ states (small dashed blue arrows) are carried over to the $^3$P$_0$ states due to the connection through the laser. Additionally, the $^3$P$_0$ states are shifted by the laser frequency $\tilde{\omega}$ (dashed orange arrow). The dotted black arrows show the path taken by the graph search algorithm. Note that this path changes, \eg{}, if the maximum detuning $\Delta_\mathrm{max}$ is large enough to bring the laser in resonance with the $F=\nicefrac{3}{2}\rightarrow\, F^\prime=\nicefrac{1}{2}$ transition.\label{fig:transform}}
\end{figure}

To simulate the coherent dynamics of the system, a Hamiltonian matrix is created in the basis of the atom, spanned by the user-specified states $\ket{i}$. The starting point is the time-dependent Hamiltonian
\begin{align}
H = \sum\limits_i \sigma_{ii}\,\hbar\omega_i + \sum\limits_k\sum\limits_{\substack{i,j\\i\neq j}}\sigma_{ij}\,\hbar\Omega_{ij}^k\cos(\tilde{\omega}_k t),
\label{eq:hamiltonian-al}
\end{align}
\noindent where $\sigma_{ij} \coloneqq \ket{i}\bra{j}$, $\omega = 2\pi\nu$ are circular frequencies and $\Omega_{ij}^k$ are the Rabi frequencies generated by laser $k$. This Hamiltonian describes the interaction of an atom with multiple classical laser fields. In this form, every laser drives all atomic \mbox{transitions $\sigma_{ij}$}. The \qsim{} module identifies the transitions that fulfill the electric-dipole-transition rules and calculates the Rabi frequencies from the parameters of the lasers and the properties of the involved atomic states. The full set of conditions for a transition $\sigma_{ij}$ to be considered is given by
\begin{align}
\label{eq:condition0}
|J_j - J_i| &\leq 1,\quad 0\nleftrightarrow 0\\
|F_j - F_i| &\leq 1,\quad 0\nleftrightarrow 0\\
|m_j - m_i| &\leq 1,\quad 0\nleftrightarrow 0\quad\text{if}\quad\Delta J,F = 0 \\
\Gamma_{\!ij} &> 0\\
I_k &> 0\\
(\vec{q}^{\,k})_{m_j - m_i} &\neq 0,\quad \nu_j > \nu_i\\
\left||\nu_j - \nu_i| - \tilde{\nu}_k\right| &\leq \Delta_\mathrm{max},
\label{eq:condition1}
\end{align}
with the Einstein $A_{ji}$ coefficient and a user-defined cutoff frequency $\Delta_\mathrm{max}$ to ignore fast oscillations. Since the Hamiltonian still depends on the fast oscillating terms $\cos(\tilde{\omega}_k t)$, it is transformed into an interaction picture and the rotating-wave approximation is applied to frequencies of the order of $2\tilde{\omega}_k$. The transformation is constructed to produce a time-independent Hamiltonian whenever possible to avoid complex exponentials for faster computation. This is possible if and only if two or more lasers do not form loops in the multigraph generated by connecting the atomic states with the lasers. Since the transformation depends on the particular graph, which is only determined at runtime, no closed expression for the transformed Hamiltonian can be specified. The graph search algorithm that is used for the transformation is described in Fig.\,\ref{fig:transform} based on a laser-driven $^3$S$_1\rightarrow\,^3$P$_0$ transition in a system with nuclear spin $\nicefrac{1}{2}$. The transformed time-independent Hamiltonian of this example is given by \cite{Manzano.2020}
\begin{align}
H^\prime &\coloneqq U^\dagger HU + i\hbar\frac{\mathrm{d}U^\dagger}{\mathrm{d}t}U \nonumber\\
&= \hbar(\omega_2 - \omega_0 - \tilde{\omega})(\hat{\sigma}_{22} + \hat{\sigma}_{33}) \nonumber\\
&\quad + \hbar\frac{\Omega}{2}(\hat{\sigma}_{02} + \hat{\sigma}_{13}) + \hbar\frac{\Omega^\ast}{2}(\hat{\sigma}_{20} + \hat{\sigma}_{31}),
\end{align}
where the indexes correspond to the states indicated in the simplified graph in Fig.\,\ref{fig:transform}, $\tilde{\omega}$ and $\Omega$ are the laser and Rabi frequencies, respectively, $U$ is the unitary transformation defined in Fig.\,\ref{fig:transform}, and $\omega_0 = \omega_1$ as well as $\omega_2 = \omega_3$ due to the absence of an external magnetic field. When the time-dependence cannot be eliminated completely in the presence of multiple lasers, the off-diagonal elements of the Hamiltonian will contain time-dependent Rabi frequencies, which oscillate with the frequency differences between the lasers. This is automatically detected by the algorithm and the Hamiltonian is adjusted accordingly.

\subsection{Lindblad Master equation}
\label{sec:lindblad master equation}

The complete coherent and dissipative population dynamics of the atomic states in the laser fields can be determined by solving the Lindblad master equation. It can be written as \cite{Lindblad.1976, Carmichael.1999, Manzano.2020}
\begin{align}
\label{eq:master}
\frac{\partial\rho}{\partial t} &= -\frac{i}{\hbar}[H, \rho] + \sum\limits_{i,j} \Gamma_{\!ij}\mathcal{D}[\sigma_{ij}]\rho\\
\mathcal{D}[\sigma]\rho &\coloneqq \sigma\rho\sigma^\dagger - \frac{1}{2}(\sigma^\dagger\sigma\rho + \rho\sigma^\dagger\sigma),\label{eq:lindblad-operator}
\end{align}
where the coherent dynamics are described by the first term and the coupling to the vacuum by the second term. Note that $\Gamma_{\!ij} \neq 0 \Rightarrow \Gamma_{\!ji} = 0$ as spontaneous decay from $\ket{i}$ to $\ket{j}$ only occurs if $\omega_i > \omega_j$. In \qsim{}, the differential Eq.\,\eqref{eq:master} is solved numerically in matrix form using the \cpp{} libraries \texttt{Boost.Numeric.Odeint} and \texttt{Eigen}. Large parameter spaces or numbers of samples can be explored by specifying frequency detunings, velocities (Doppler shifts) or initial density matrices (state populations) as \np{} arrays, which are addressed in parallel by the \cpp{} backend. Hence, a user is neither required to write any loops in Python nor to implement parallel computing. The performance of the master equation solvers in \qs{} and \qu{} for a single parameter sample are comparable and mainly depend on the chosen integration algorithm, integration steps and error tolerances.\\
A MC master equation solver is available, following the approach described in \cite{Dum.1992}, which is also implemented in \qu{} \cite{Johansson.2012}. The basic idea is to solve the Schr\"{o}dinger equation with a modified non-hermitian Hamiltonian that leaks population into the environment according to the dissipation rates $\Gamma_{\!ij}$. Whenever the norm of a simulated state vector $\ket{\psi_s}$, where $s$ numerates the samples, falls below a randomly generated fraction of unity, the state vector ``collapses'', destroying all superpositions and resetting the system to the collapsed state. Averaging the resulting state vectors for $n\rightarrow\infty$ samples yields the density matrix described by the master equation
\begin{align}
\rho_{ij} = \lim\limits_{n\rightarrow\infty}\frac{1}{n}\sum\limits_{s=1}^n \braket{i|\psi_s}\braket{\psi_s|j}.
\label{eq:mc density}
\end{align}
The MC method is particularly useful for large atomic systems since its complexity only increases linearly with the number of atomic states as opposed to the quadratic growth of the master equation. In \qsim{}, the MC master equation additionally can be used to simulate the motional degrees of freedom of the atom. Whenever the atom spontaneously decays, the photon momenta of the lasers leading from the initial to the excited state and the recoil from the spontaneous decay are added to the momentum of the atom. Note that in this method, the momentum of the atom only changes upon spontaneous decays but not during any coherent interaction.

\subsection{Rate equations}
\label{sec:rate equations}

A simplification of the master equation are the rate equations which omit all coherent dynamics. In laser spectroscopy, the use of rate equations is often sufficient due to how precise the state populations need to be known or due to the width of the velocity distribution of the atomic ensemble, which can average out coherent effects. The rate equations implemented in \qsim{} are given by
\begin{align}
\frac{\partial\rho_{ii}}{\partial t} = \sum\limits_j \left[ \left(\sum\limits_k R_{ij}^k\right)(\rho_{jj} - \rho_{ii}) + \Gamma_{\!ij}\,\rho_{jj} - \Gamma_{\!ji}\,\rho_{ii}\right],
\label{eq:rate equations}
\end{align}
where $R_{ij}^k$ are the stimulated emission/absorption rates for laser $k$, which depend on the frequency, intensity and polarization of the laser, $\Gamma_{\!ij}$ are the spontaneous decay rates from Eq.\,\eqref{eq:master} and $\rho_{ij} = 0$ for $i\neq j$.

\subsection{Differential scattering rate}
\label{sec:differential scattering rate}

From the density matrices determined through Eq.\,\eqref{eq:master}, \eqref{eq:mc density} and \eqref{eq:rate equations}, a differential scattering rate into the solid angle $\Omega$ can be calculated that, in case of the coherent approaches, also considers quantum interference (QI) effects. The differential scattering rate is given by \cite{Drake.2023}
\begin{align}
&\frac{\mathrm{d}\Gamma}{\mathrm{d}\Omega}(\vec{k}_\mathrm{sc}) \coloneqq \mathrm{Tr}\left(\rho D(\vec{k}_\mathrm{sc})\right)\nonumber\\
&\quad = \frac{1}{4\pi}\sum\limits_{\vec{q}\in\mathcal{P}(\vec{k}_\mathrm{sc})}\ \sum\limits_{f\in\mathcal{F}}\ \sum\limits_{i,j\in\mathcal{I}_f}\sqrt{\Gamma_{\!\!fi}\Gamma_{\!\!f\!j}}\,\,\frac{(\vec{q}\cdot \vec{d}_{fi})(\vec{d}_{f\!j}^\ast\cdot\vec{q}^\ast)}{|\vec{d}_{fi}||\vec{d}_{f\!j}^\ast|}\,\rho_{ij}^\ast,
\label{eq:scattering rate}
\end{align}
where $D(\vec{k}_\mathrm{sc})$ is a detection operator, $\mathcal{P}(\vec{k}_\mathrm{sc})$ is a basis set of normalized polarization vectors for scattered photons with direction $\vec{k}_\mathrm{sc}$, $(\mathcal{I}_f)_{f\in\mathcal{F}}$ and $\mathcal{F}$ are sets of excited and final states between which spontaneous decay can occur and $\vec{d}$ are the dipole transition strengths, see Eq.\,\eqref{eq:dipole operator}. Limiting the three sums to detectable spontaneous decay and integrating over the solid angle of detection yields a photon detection rate.

\begin{figure}[!t]
\centering
\includegraphics[width=\linewidth]{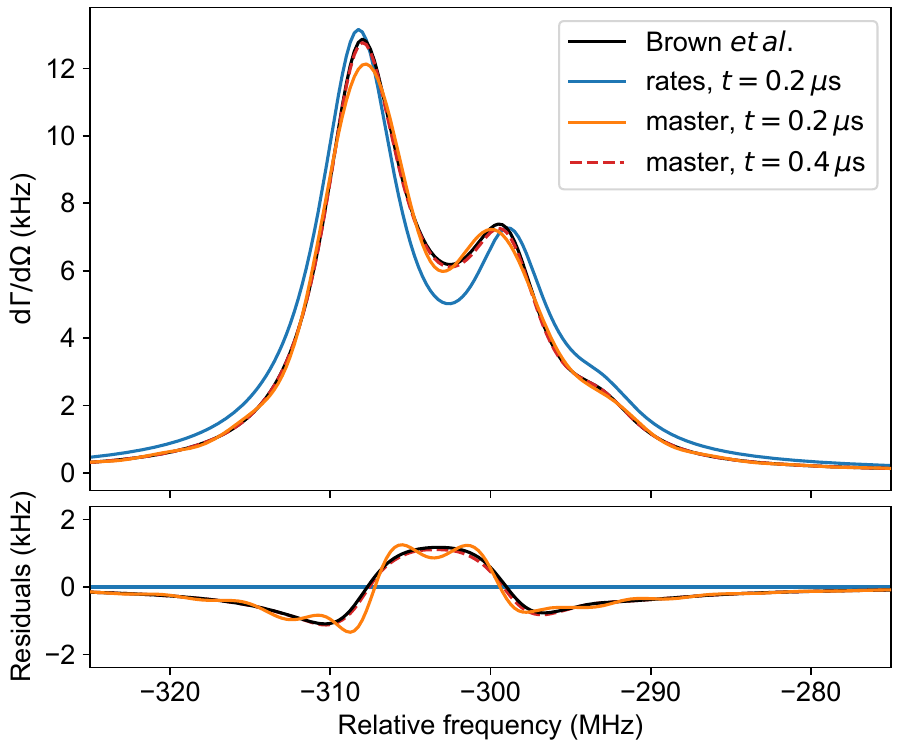}
\caption{Simulated differential scattering rate of the $^2$S$_{1/2}(F=2)\rightarrow\,^2$P$_{3/2}$ transitions in $^7$Li in the direction of the linear polarization vector of a spectroscopy laser with an intensity of $1\,\upmu\mathrm{W}/\mathrm{mm}^2$. The perturbative scattering rate, which includes quantum interference (QI) effects derived by Brown\etal{} \cite{Brown.2013} (black) is compared to the solutions of the rate (blue) and master (orange) equations after an integration time of $0.2\,\upmu$s. While the rate equations yield a sum of Lorentzians with averaged out coherent and QI effects, the master equation approach shows the intensity modulations of the coherent Rabi oscillations and converges to the perturbative QI approach for later times (red dashed). A small difference comes from optical population transfer into the $F = 1$ ground state. This plot reproduces Fig.\,2 of \cite{Brown.2013}. The shown data was generated using \mbox{Code Listing \ref{code:simulate}}.\label{fig:simulate}}
\end{figure}

Equation \eqref{eq:scattering rate} reproduces the perturbative differential scattering rate derived by Brown\etal{}~\cite{Brown.2013}, see Eq.\,\eqref{eq:scattering rate pert}, that considers QI effects in closed electronic transitions in the case of small laser intensities, such that $\Omega_{ij} \ll \Gamma_{\!ij}$, linear polarization and equilibrium population. These conditions are rarely met exactly in the experiment as in most real atomic systems, especially if they have a hyperfine structure, optical pumping and population transfer will occur already within short interaction times and laser powers on the $\upmu\mathrm{W}/\mathrm{mm}^2$ level. A comparison of differential scattering rates of the $^2$S$_{1/2}\rightarrow\,^2$P$_{3/2}$ transition in $^7$Li is shown in Fig.\,\ref{fig:simulate}. The scattering rates were determined with Eq.\,\eqref{eq:scattering rate} using the master and the rate equation approaches as well as with the perturbative scattering rate from \cite{Brown.2013} for detection at $0^\circ$ angle relative to the polarization vector of absorbed photons. A code example that produces the data shown in Fig.\,\ref{fig:simulate} is given in Code Listing \ref{code:simulate}, using data from \cite{Brown.2013, NIST}. The shown plot reproduces Fig.\,2 from \cite{Brown.2013} and additionally shows the solution of the master equation approach after an integration time of $0.2$ and $0.4\,\upmu$s, demonstrating the need of the non-perturbative approach if the system equilibrium is not reached yet.

\section{Summary and outlook}
\label{sec:summary and outlook}

The Python package \qs{} provides a variety of classes and functions to facilitate the analysis and simulation workflow in laser spectroscopy experiments. In this article, a multidimensional linear regression algorithm, modular fit models and light-matter simulations are introduced. The maximum-likelihood fit of a straight line in multiple dimensions constitutes a generalization of the linear regression algorithm introduced by York\etal{}~\cite{York.2004}. The system of modular fit models provides a basis for combining and linking different fit models, parameters and data sets and can be easily expanded with new models. The simulation module for coherent light-matter interactions offers an intuitive and user-friendly interface without the need to specify a Hamiltonian or other complicated mathematical objects.\\
The \qs{} package is built to be as general as possible within the given set of physical conditions. An expansion of \qs{} with new features and further generalization for a larger scope of applications is planned. For example, the simulation module can be used to define a new lineshape model for fitting, which can account for the population dynamics of an atom in a laser field. The simulation module itself can be expanded, \eg{}, by accounting for finite laser linewidths, adding molecular systems or allowing higher-order multipole transitions or time-dependent velocities and laser parameters.

\section*{Acknowledgements}

We thank K. K\"onig, B. Maa\ss, E. Matthews, K. Mohr, L. Renth and J. Spahn for testing the \qs{} package and their valuable feedback. We acknowledge support by the Deutsche Forschungsgemeinschaft (DFG, German Research Foundation) under - Projektnummer 279384907 - SFB 1245, and the German Federal Ministry of Education and Research 
BMBF under Contract Nos. 05P21RDFN1 and 05P21RDCI1. P.\,M. acknowledges support from \mbox{HGS-HIRE}.

\appendix

\section{Gradient and Hessian of the negated log-likelihood function}
\label{sec:gradient and hessian of the negated log-likelihood function}

For the minimization of the negative log-likelihood function given by Eq.\,\eqref{eq:log-likelihood}, its gradient vector and Hessian are calculated analytically. Both are used in the Newton conjugate gradient (Newton-CG) method \cite{scipy, Nocedal.2006} and the Hessian is also used to estimate the covariance matrix of the resulting parameters. We define
\begin{align}
\tilde{a}_{ij}\coloneqq \left(\Sigma_i^{-1}\vec{a}_i\right)_j,\qquad \tilde{b}_{ij}\coloneqq \left(\Sigma_i^{-1}\vec{b}\right)_j.
\end{align}
Then the components of the gradient vector of Eq.\,\eqref{eq:log-likelihood} are given by
\begin{align}
\frac{\partial}{\partial a_j}\mathcal{L} &= \sum\limits_{i\in\mathcal{I}}\left[\tilde{a}_{ij} - \frac{t_i}{\sigma_i^2}\tilde{b}_{ij}\right] \\
\frac{\partial}{\partial b_j}\mathcal{L} &= \sum\limits_{i\in\mathcal{I}}\left[\frac{t_i}{\sigma_i^2}\tilde{a}_{ij} - \frac{t_i^2}{\sigma_i^4}\tilde{b}_{ij}\right] ,
\end{align}
and the components of the Hessian are given by
\begin{align}
\frac{\partial^2}{\partial a_j\, \partial a_l}\mathcal{L} &= \sum\limits_{i\in\mathcal{I}}\left[ \left(\Sigma_i^{-1}\right)_{jl} - \frac{\tilde{b}_{ij}\tilde{b}_{il}}{\sigma_i^2}\right]\\
\frac{\partial^2}{\partial b_j\,\partial b_l}\mathcal{L} &= \sum\limits_{i\in\mathcal{I}} 
\left[
\frac{t_i^2}{\sigma_i^4}\left(\Sigma_i^{-1}\right)_{jl} - \frac{1}{\sigma_i^2}\tilde{a}_{ij}\tilde{a}_{il} - 4\frac{t_i^2}{\sigma_i^6}\tilde{b}_{ij}\tilde{b}_{il}
\right. \nonumber\\
&\left. \quad\qquad + 2\frac{t_i}{\sigma_i^4}\left(\tilde{a}_{ij}\tilde{b}_{il} + \tilde{b}_{ij}\tilde{a}_{il}\right)\right]\\
\frac{\partial^2}{\partial a_j\,\partial b_l}\mathcal{L} &= \sum\limits_{i\in\mathcal{I}}\left[-\frac{t_i^2}{\sigma_i^2}\left(\Sigma_i^{-1}\right)_{jl} + 2\frac{t_i}{\sigma_i^4}\tilde{b}_{ij}\tilde{b}_{il} - \frac{1}{\sigma_i^2}\tilde{a}_{ij}\tilde{b}_{il}\right].
\end{align}

\section{Definitions for \qsim{}}
\label{sec:definitions for qspec.simulate}

The following definitions are used to calculate light-matter interactions in \qsim{} \cite{Hilborn.1982, Brown.2013}
\begin{align}
\nu_i &= \nu_{J_i} + \nu^{F_i}_{(\gamma_i J_i)(\gamma_i J_i)} - m_{F_i}g_{F_i}\frac{\mu_\mathrm{B}}{h}\mathcal{B}\\[1ex]
g_F &= g_J\,\frac{F(F + 1) + J(J + 1) - I(I + 1)}{2F(F + 1)} \nonumber\\
&\quad + g_I\,\frac{\mu_\mathrm{N}}{\mu_\mathrm{B}} \frac{F(F + 1) - J(J + 1) + I(I + 1)}{2F(F + 1)}\\[1ex]
g_J &= -\frac{J(J + 1) + L(L + 1) - S(S + 1)}{2J(J + 1)}\nonumber\\
&\quad+ g_s\,\frac{J(J + 1) - L(L + 1) + S(S + 1)}{2J(J + 1)},\\[3ex]
a_{ij} &= (-1)^{I + J_j + F_i + 1} \sqrt{2F_i + 1}\sqrt{2J_j + 1}\nonumber\\
&\quad\times\braket{F_i m_i; 1\, m_j - m_i|F_j m_j}\begin{Bmatrix}J_j & J_i & 1\\F_i & F_j & I\end{Bmatrix}\\
\tilde{\Gamma}_{ij} &= \begin{cases}A_{ji} & \,\nu_i < \nu_j\\A_{ij} & \, \text{else}\end{cases}\\
\Gamma_{ij} &= a_{ij}^2\,A_{ji}\\
d_{ij} &= \sqrt{\frac{3\varepsilon_0 c^3 \hbar\tilde{\Gamma}_{\!ij}}{8\pi^2|\nu_j - \nu_i|^3}}\,a_{ij}\label{eq:dipole operator}\\
M_{ij}^k &= \begin{cases} 1 & \,\text{Eq.\,\eqref{eq:condition0} - \eqref{eq:condition1} fulfilled}\\ 0 & \, \text{else}\end{cases}\\
\Omega_{ij}^k &= \sqrt{\frac{2I_k}{\varepsilon_0 c}}\frac{d_{ij}(\vec{q}^{\,k})_{m_j - m_i}}{\hbar}M_{ij}^k\\
R_{ij}^k &= \frac{|\Omega_{ij}^k|^2\,\tilde{\Gamma}_{\!\!ji}}{(2\pi)^2(\tilde{\nu}_k - |\nu_j - \nu_i|)^2 + \frac{1}{4}\tilde{\Gamma}_{\!\!ji}^2},
\end{align}
where $i$ and $j$ enumerate individual quantum states, $\nu_i$ are the eigen frequencies, $m_i$ are the $z$-projection quantum numbers of the total angular momentum quantum numbers $F_i$, $\mathcal{B}$ is the magnetic flux density, $g$ are the Land\'e factors, $\mu_\mathrm{N}$ and $\mu_\mathrm{B}$ are the nuclear and Bohr magneton, respectively, $\braket{\cdot\cdot\cdot\cdot|\cdot\cdot}$ are the Clebsch-Gordan coefficients, \mbox{$\lbrace :\,:\,:\rbrace$} are the Wigner-$6j$ symbols, $\vec{q}^{\,k}$ is a normalized complex laser polarization vector of laser $k$, $A_{ji}$ are Einstein coefficients, $I_k$ are the laser intensities, $\varepsilon_0$ is the vacuum permittivity and $c$ is the vacuum speed of light.

\newpage
\section{Class diagrams of \qmod{} and \qsim{}}
\label{sec:class diagrams of qspec.models and qspec.simulate}

\begin{strip}
\centering
\captionsetup{type=figure}
\includegraphics[width=0.75\linewidth]{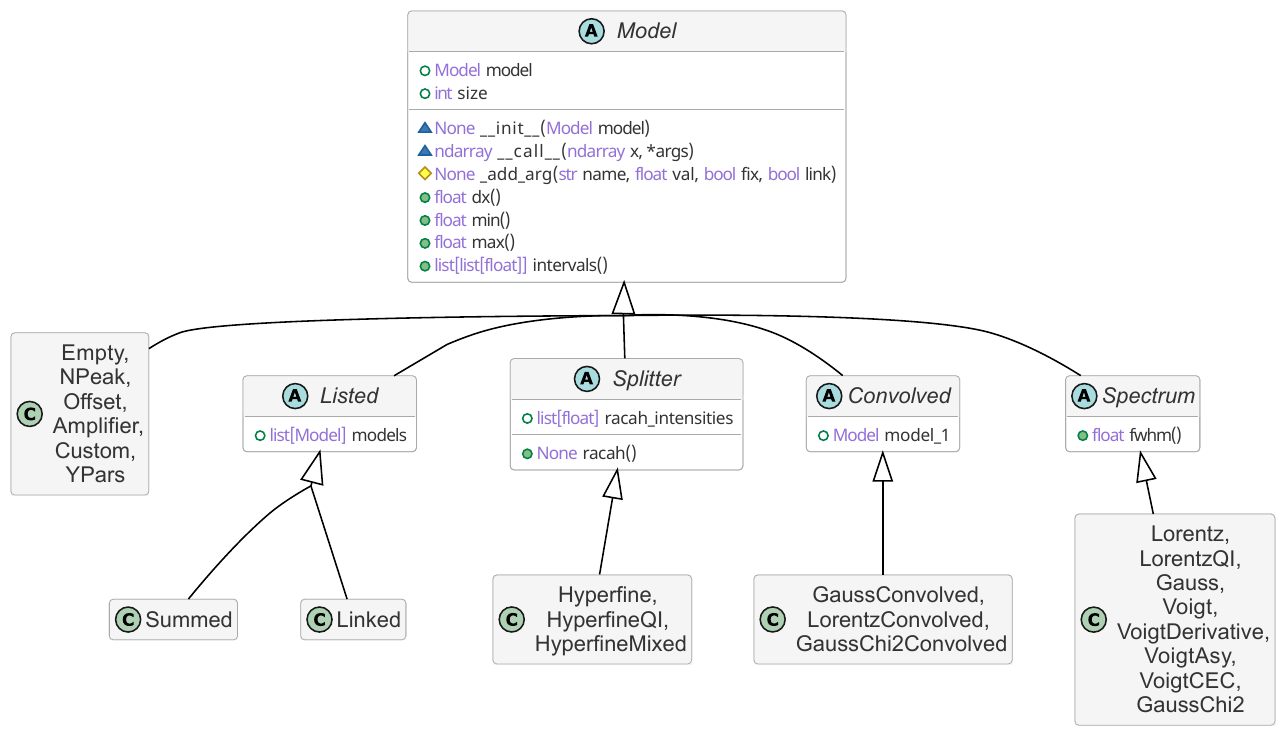}
\captionof{figure}{Class inheritance diagram of \qmod{}. All models inherit from the \texttt{Model} class. The abstract \texttt{Listed} class enables the combination of models to link parameters or sum models. The \texttt{Splitter} class is an abstract class for Hyperfine structure models. \texttt{Convolved} allows numerical convolution of two models and \texttt{Spectrum} is a generic lineshape/peak model.}\label{fig:models-uml}
\vfill
\includegraphics[width=0.75\linewidth]{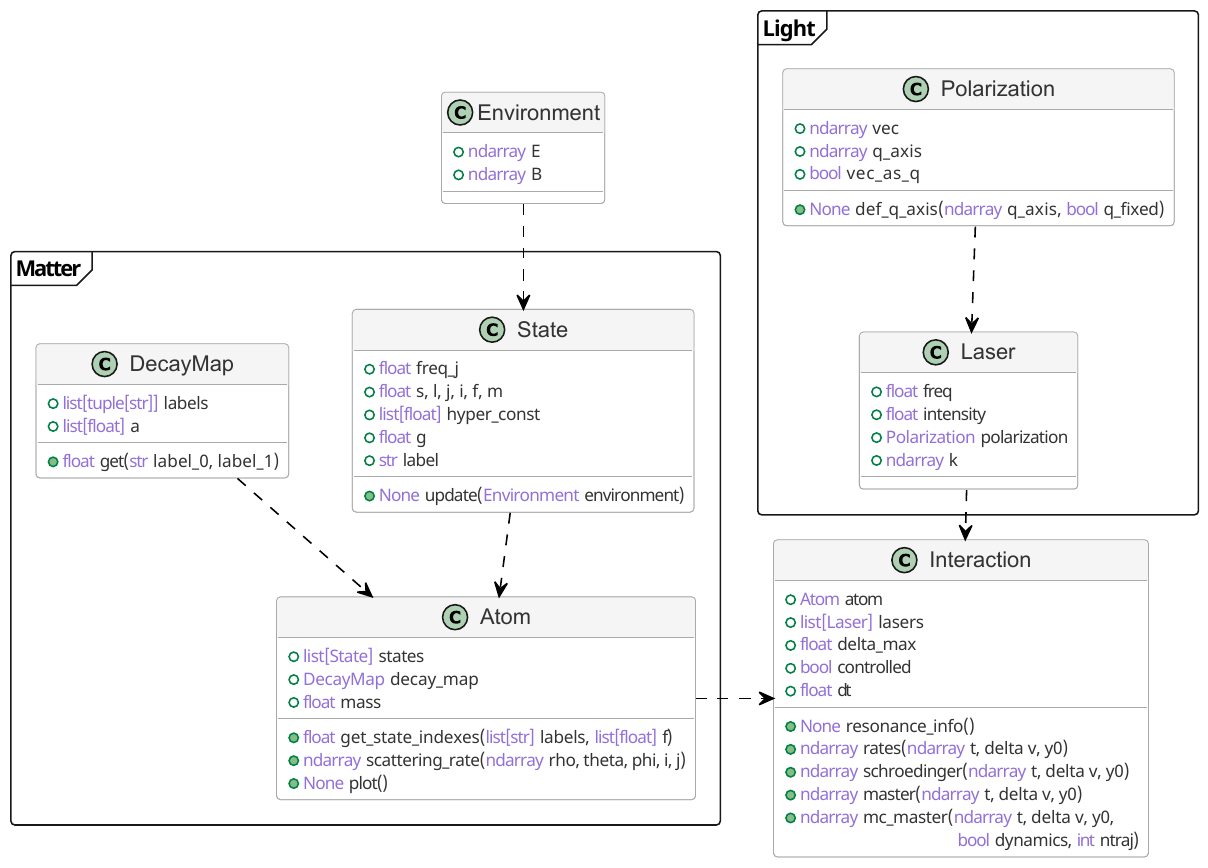}
\captionof{figure}{Class diagram of \qsim{}. A \texttt{State} represents a single quantum mechanical state. A \texttt{DecayMap} object links multiple states with the same user-defined labels through spontaneous decay. A list of all states and the \texttt{DecayMap} object form an \texttt{Atom}. A \texttt{Laser} consists of a frequency, an intensity, a \texttt{Polarization} and a directional vector. The \texttt{Atom} and a list of lasers is combined in an \texttt{Interaction}. Rate, Schr\"odinger, master and a Monte-Carlo (MC) master equation solver can be accessed directly from the \texttt{Interaction}. An \texttt{Environment} can be defined to alter the quantum state energies, but is not fully implemented yet and currently only supports linear Zeeman shifts. Note that the list of shown member functions is not complete for better clarity.\label{fig:simulate-uml}}
\end{strip}

\newpage~\newpage~\newpage
\section{Code examples of \qs{}}
\label{sec:code examples of qspec}

\begin{strip}
\begin{python}[caption={Example code to perform a 2d King-plot analysis with Ca$^+$ isotopes. In this example two transitions, named D1 and D2 line, are plotted against each other for the stable Ca$^+$ isotopes \cite{Mueller.2020}. Transition frequencies are specified as absolute values (line 14-21) so that \qs{} can calculate isotope shifts at runtime, which facilitates reassignments of reference isotopes (line 28-29). The fitted parameters of the straight line are used to determine the isotope shifts of the D1 line in $^{50,52}$Ca$^+$ by inserting the isotope shifts of the D2 line (line 42-43) \cite{GarciaRuiz.2016}. Finally the results are printed to the console output (line 45-47). Masses are taken from the Atomic Mass Evaluation (AME) 2020 \cite{Wang.2021}. Note that this example can be extended to an $n$-dimensional fit simply by adding more columns/observables to \texttt{x\_abs}. If absolute values are unknown, the isotope shifts have to be passed to the \texttt{king.fit} routine in the same format as \texttt{x\_abs} and need to concur with the specified lists of (reference) isotopes.\label{code:king-plot}}]
import qspec as qs

# The mass numbers of the Ca isotopes.
a = [40, 42, 43, 44, 46, 48, 50, 52]

# The masses of the isotopes (u, AME 2020).
m = [(39.962590850,   22e-9), (41.958617780, 159e-9),  # 40Ca, 42Ca
     (42.958766381,  244e-9), (43.955481489, 348e-9),  # 43Ca, 44Ca
     (45.953687726, 2398e-9), (47.952522654,  18e-9),  # 46Ca, 48Ca
     (49.957499215, 1.7e-6 ), (51.963213646, 720e-9)]  # 50Ca, 52Ca

# Use absolute values given in the shape (#isotopes, #observables, 2).
# Frequencies for the (D1, D2) lines (MHz).
x_abs = [[(755222765.66, 0.10), (761905012.53, 0.11)],  # 40Ca
         [(755223191.15, 0.10), (761905438.57, 0.10)],  # 42Ca
         [(755223443.57, 0.30), (761905691.89, 0.17)],  # 43Ca
         [(755223614.66, 0.10), (761905862.62, 0.09)],  # 44Ca
         [(755224063.27, 0.33), (761906311.60, 0.57)],  # 46Ca
         [(755224471.12, 0.10), (761906720.11, 0.11)],  # 48Ca
         [(        0.  , 0.  ), (        0.  , 0.  )],  # 50Ca
         [(        0.  , 0.  ), (        0.  , 0.  )]]  # 52Ca

# Construct a King object. Optionally specify 'x_abs' here
# to omit isotope shifts when fitting. 20 electron masses are subtracted
# to perform the King plot analysis with the nuclear masses.
king = qs.King(a=a, m=m, x_abs=x_abs, subtract_electrons=20)

a_fit = [42, 43, 44, 46, 48]  # Choose the isotopes to fit.
a_ref = [40, 48, 42, 40, 44]  # Choose individual reference isotopes.

# Do a simple 2d King plot. The 'mode' keyword is only used for the axis labels.
popt, pcov = king.fit(a_fit, a_ref, mode='shifts')
# >>> f(x) = (177.3 u MHz) + 1.00068 * x

a_unknown = [50, 52]  # Specify the unknown isotopes
a_unknown_ref = [40, 40]  # and their references.

# Specify the isotope shifts of the D2 line.
y = [(1969.2, 5.6), (2219.2, 7.0)]

# Calculate the isotope shifts of the D1 line and their covariances.
x, cov, cov_stat = king.get_unmodified(
    a_unknown, a_unknown_ref, y, axis=1, show=True, mode='shifts')

for iso, c in zip(a_unknown, cov):
    qs.printh(f'\n{iso}Ca+:')  # Print colored headline.
    qs.print_cov(c)  # Print color-coded covariance matrix.
\end{python}
\end{strip}

\newpage~\newpage~\newpage
\begin{strip}
\begin{python}[caption={Example code to produce the data shown in Fig.\,\ref{fig:simulate}. Lists of the magnetic substates of the ground (\texttt{s}) and the excited state (\texttt{p}) are created (line 10-13), connected in a \texttt{DecayMap} (line 15) and used to define a $^7$Li atom (line 16). A laser is defined with an intensity of $1\,\upmu \mathrm{W}/\mathrm{mm}^2$ and $\pi$-polarized light (line 18-20). An \texttt{Interaction} object combines the atom and a list of lasers (line 22). The rate (line 29) and master (line 32, 35) equation solvers are called directly from the \texttt{Interaction} object for an array of frequency detunings \texttt{delta}. The scattering rate for the different detunings defined in Eq.\,\eqref{eq:scattering rate} is calculated directly from the array of density matrices \texttt{rho} or state populations \texttt{n} (line 30, 33, 36). An extended version of the perturbative scattering rate derived by Brown\etal{}~\cite{Brown.2013} is calculated using a \texttt{ScatteringRate} object (line 38-39). Transition frequencies and hyperfine structure constants are taken or calculated from the data in \cite{Brown.2013}. The Einstein coefficient is taken from \cite{NIST}.\label{code:simulate}}]
import numpy as np
import qspec.simulate as sim

f_sp = 446810183.163  # Transition frequency (MHz)
a_sp = 36.891  # Einstein coefficient (rad MHz)

s_hyper = [401.75825]  # HFS constants (MHz)
p_hyper = [-3.055038, -0.29670]

s = sim.construct_electronic_state(
    0., s=0.5, l=0, j=0.5, i=1.5, hyper_const=s_hyper, label='s')
p = sim.construct_electronic_state(
    f_sp, s=0.5, l=1, j=1.5, i=1.5, hyper_const=p_hyper, label='p')

decay = sim.DecayMap(labels=[('s', 'p')], a=[a_sp])
li7 = sim.Atom(s + p, decay)

intensity = 1.  # uW /mm**2
polarization = sim.Polarization([0, 1, 0])  # Linear polarization
laser = sim.Laser(f_sp, intensity, polarization)

inter = sim.Interaction(li7, [laser])
inter.controlled = True  # Error controlled integrator

t = 0.2  # Integration time (us)
delta = np.linspace(-325, -275, 201)  # Frequency detunings (MHz)
theta, phi = 0., 0.  # Angles from z-axis in x- and y-direction (rad)

n = inter.rates(t, delta)  # Rate equations, 0.2 us
y_rates = li7.scattering_rate(n, theta, phi, as_density_matrix=False)

rho = inter.master(t, delta)  # Master equation, 0.2 us
y_master = li7.scattering_rate(rho, theta, phi)

rho = inter.master(0.4, delta)  # Master equation, 0.4 us
y4_master = li7.scattering_rate(rho, theta, phi)

sr = sim.ScatteringRate(li7, laser=laser)
y_brown = sr.generate_y(delta, theta, phi)[:, 0, 0]  # Brown et al.
\end{python}
\end{strip}

\bibliographystyle{elsarticle-num}
\bibliography{bibliography}

\end{document}